\documentclass[12pt,aps,floats]{revtex4}
\usepackage{graphicx}
\usepackage{epsfig}
\usepackage{amssymb}
\usepackage{flushend}\usepackage{amsmath}\usepackage{amssymb}
\usepackage{indentfirst}
\usepackage{bm}

\newskip\humongous \humongous=0pt plus 1000pt minus 1000pt

\relax

\begin{document}

\title{Quantum tunneling from high dimensional G\"{o}del black hole}
\bigskip
\author{Hui-Ling Li$^{a,b}\footnote{Electronic address: LHL51759@126.com}$, Zhong-Wen Feng$^{a}$, Xiao-Tao Zu$^{a}$}
\affiliation{$^{a}$ School of Physical Electronics, University of Electronic Science and
Technology of China, Chengdu, 610054, People's Republic of China \\ $^{b}$ College of Physics Science and Technology, \\
Shenyang Normal University, Shenyang 110034, People's Republic of China}

\begin{abstract}
\textbf{Abstract:}
Considering quantum gravity effect, we investigate the quantum tunneling from high dimensional Kerr-G\"{o}del black hole using generalized Dirac equation. As a result, revised tunneling probability is obtained, and the corrected Hawking temperature is also presented.


\end{abstract}

\maketitle

\section{Introduction}

Since Hawking made a striking discovery that a black hole can radiate particles from its event horizon with a black body spectrum and the temperature is proportional to its surface gravity \cite{1}, a considerable amount of work has been done relating to static, stationary and non-stationary black holes' Hawking radiation \cite{2,3,4,5,6,7,8,9,10,11,12,13,14,15,16,17,18}. Recently, incorporating quantum gravity effects into black hole's physics, generalized uncertainty principle (GUP) with a minimal measurable length has attracted a lot of interest \cite{19,20,21,22,23,24,25,26}. A minimal observable length arises naturally when incorporating gravity into quantum field, which also leads to a modification of original uncertainty principle. This modification now called GUP. GUP produces some very interesting and novel implications and influences the well know traditional Hawking radiation and revises a serise of results dramatically. Consider a GUP with a minimal observable length, applying elegant Parikh-Wilczek tunneling method, K. Nozari et al. \cite{23} have succeeded in presenting scalar particle's quantum tunneling from a schwarzchild black hole. At the same time, in this framework, this method has been extended to noncommutative black holes \cite{24}. Latter, considering a GUP with natural cutoffs as a minimal length, a minimal momentum and a maximal momentum, K. Nozari et al. \cite{27} discussed on quantum tunneling of massless particles' acrossing black hole horizon and remarked on the role played by these natural cutoffs on the tunneling probability. On the other hand, a GUP with a minimal measurable length also means a correction to Dirac equation in curved spacetime. Very recently, adopting Kerner and Mann's Hamilton-Jacobi anatz and modified Dirac equation based on GUP, Chen et al. \cite{28,29,30} addressed the fermions tunneling from black holes and black rings, and discussed on the issue about black hole evaporation and remnants. However, as for higher dimensional black holes, the issue of quantum corrections to fermion's Hawking radiation is comparatively less discussed.

Higher dimensional black holes and their properties have attracted considerable attention \cite{31,32,33,34,35}, in particular, with the development of the conformal field theory (CFT) correspondence, allowing statement to be made about four dimensional quantum field theory using solutions of the Einstein equation constant in five dimensions, and with the advent of brane-world theories \cite{36,37}, raising the possibility of direct observation of Hawking radiation and of as probes of large spatial extra dimensional in future high energy colliders \cite{38,39}. On the other hand, a realistic black hole must be localized inside the cosmological background, and a natural consideration for it is our universe. The G\"{o}del universe is an exact solution to Einstein's equation in the presence of a cosmological constant and homogeneous pressureless matter. In recent years, there has been great interest in studying G\"{o}del-type solutions to five-dimensional supergravity \cite{40,41,42,43,44,45}. These developments motivate us to explore five dimensional G\"{o}del black hole's quantum correction to fermion tunneling via modified Dirac equation based on the GUP.

\section{A brief review of the modified Dirac equation}

In a curved background, the Dirac equation can be written as
\begin{eqnarray} \label{1}
i\gamma^{\mu}\left(\partial_{\mu}+\Omega_{\mu}+\frac{i}{\hbar}eA_{\mu}\right)\psi+\frac{\mu}{\hbar}\psi=0,
\end{eqnarray}
where
\begin{eqnarray} \label{2}
\Omega_{\mu}=\frac{i}{2}\omega_{\mu}^{\phantom{\mu}ab}\Sigma_{ab},\qquad
\Sigma_{ab}=\frac{i}{4}[\gamma^{a},\gamma^{b}],\qquad
\left\{\gamma^{a},\gamma^{b}\right\}=2\eta^{ab},\nonumber\\
\omega_{\mu\phantom{a}b}^{\phantom{\mu}a}=e_{\nu}^{\phantom{\nu}a}e^{\lambda}_{\phantom{\lambda}b}\Gamma_{\mu\lambda}^{\nu}
-e^{\lambda}_{\phantom{\lambda}b}\partial_{\mu}e_{\lambda}^{\phantom{\lambda}a},\qquad
\gamma^{\mu}=e^{\mu}_{\phantom{\mu}a}\gamma^{a},\qquad \left\{\gamma^{\mu},\gamma^{\nu}\right\}=2g^{\mu\nu}.
\end{eqnarray}
The modified Dirac equation is suggested on the basis of a minimal observable length. In the framework of generalized uncertainty principle
(GUP), this length is obtained. In one dimension, the GUP is given by
\begin{eqnarray} \label{3}
\triangle x\triangle P\geqslant \frac{\hbar}{2}[1+\beta(\triangle P)^{2}],
\end{eqnarray}
where the value of $\beta=\beta_{0}l_{P}^{2}/\hbar^{2}$ is very small, the dimensionless parameter $\beta_{0}$ satisfies $\beta_{0}<10^{-34}$ and $l_{P}$ stands for Planck length. On the basis of the corrected Heisenberg algebra $[x_{i}, P_{j}]=i\hbar \delta_{ij}[1+\beta P^{2}]$, the GUP is presented. Position operator $x_{i}$ and momentum operator $ P_{j}$ are defined as \cite{46}
\begin{eqnarray} \label{4}
x_{i}=x_{0i},\nonumber\\
P_{i}=P_{0i}(1+\beta P^{2}),
\end{eqnarray}
where $P_{0}^{2}=\Sigma P_{0j}P_{0j}$, the operators $P_{0j}$ and $x_{0i}$ satisfy the usual fundamental commutation relations $[x_{0i},P_{0j}]=i\hbar\delta_{ij}$. Other modifications may be refereed to [47-49]. Substituting Eq. (4) into Eq. (1), in a curved background, the generalized Dirac equation was presented by the revised commutation relations \cite{50}, which can be expressed as follows \cite{28,29,30}
\begin{eqnarray} \label{5}
&&\Big[i\gamma^{0}\partial_{0}+i\gamma^{i}\partial_{i}\left(1-\beta \mu^{2}\right)+i\gamma^{i}\beta \hbar^{2}\left(\partial_{j}\partial^{j}\right)\partial_{i}
+\frac{\mu}{\hbar}\left(1+\beta \hbar^{2}\partial_{j}\partial^{j}-\beta \mu^{2}\right)\nonumber\\&&+i\gamma^{\mu}\frac{i}{\hbar}eA_{\mu}\left(1+\beta \hbar^{2}\partial_{j}\partial^{j}-\beta \mu^{2}\right)+i\gamma^{\mu}\Omega_{\mu}\left(1+\beta \hbar^{2}\partial_{j}\partial^{j}-\beta \mu^{2}\right)\Big]\psi=0.
\end{eqnarray}
In the following section, using the generalized Dirac equation in Eq.$(5)$, we will focus on discuss the Kerr-G\"{o}del black hole's quantum correction to the fermion tunneling in the five-dimensional gravity background.

\section{Quantum correction to fermion tunneling}
The solution of five dimensional Kerr-G\"{o}del black hole is of the Schwarzschild-Kerr black hole embedded in a G\"{o}del universe. The metric can be expressed as \cite{45}
\begin{eqnarray} \label{6}
{\rm d}s^{2}= &-&
\!\!f(r){\rm d}t^{2}+\frac{1}{V(r)}{\rm d}r^{2}+\frac{r^{2}}{4}{\rm d}\theta^{2}-2a(r){\rm d}t\,{\rm d}\phi\nonumber\\
\!\!&+&
\!\!\left(\frac{r^{2}}{4}\sin^{2}\theta+\frac{r^{2}V(r)-4a^{2}(r)}{4f(r)}\cos^{2}\theta\right){\rm d}\psi^{2}\nonumber\\
\!\!&+&
\!\!\frac{r^{2}V(r)-4a^{2}(r)}{4f(r)}{\rm d}\phi^{2}-2a(r)\cos\theta {\rm d}t\,{\rm d}\psi\nonumber\\
\!\!&+&
\!\!2\left(\frac{r^{2}V(r)-4a^{2}(r)}{4f(r)}\cos\theta\right){\rm d}\phi
\,{\rm d}\psi,
\end{eqnarray}
where
\begin{eqnarray} \label{7}
&&f(r)=1-\frac{2m}{r^{2}},\,\,\,\,\,\,a(r)=jr^{2}+\frac{ml}{r^{2}},\nonumber\\
&&V(r)=1-\frac{2m}{r^{2}}+\frac{16j^{2}m^{2}}{r^{2}}+\frac{8jml}{r^{2}}+\frac{2ml^{2}}{r^{4}}.
\end{eqnarray}
The parameter $j$ stands for the G\"{o}del parameter and is responsible for the rotation of the spacetime. The parameter $l$ is related to the rotation of the black hole. When $j=0$ this metric reduces to the 5-dimensional Kerr black hole with the two possible rotation parameters $(l_{1},l_{2})$ of the general 5-dimensional Kerr spacetime set equal to $l$. When $l=0$ the solution becomes the Schwarzschild-G\"{o}del black hole. When the parameters $j$ and $l$ are set to zero the metric simply reduces to the 5-dimensional Schwarzschild black hole, whose mass is proportional to the parameter $m$. When $m=l=0$, the metric reduces to that of the five dimensional G\"{o}del universe.

The horizon of the black hole can be found by setting
\begin{eqnarray} \label{8}
V(r)=1-\frac{2m}{r^{2}}+\frac{16j^{2}m^{2}}{r^{2}}+\frac{8jml}{r^{2}}+\frac{2ml^{2}}{r^{4}}=0,
\end{eqnarray}
which yield outer and inter horizons,
\begin{eqnarray} \label{9}
\!\!\!\!\!\!\!\!\!r_{\pm}=\sqrt{m(1-8mj^{2}-4lj)\pm m\sqrt{(1-8mj^{2}-4lj)^{2}-\frac{2l^{2}}{m}}}.
\!\!\!\!\!\!\!\!\!\!\!\!\!\!\!\!\!\!\!\!\!\!\!\!\nonumber\\
\end{eqnarray}
At the event horizon $r_{+}$, the Hawking temperature of 5-dimensional Kerr-G\"{o}del black hole can be determined as \cite{14}
\begin{eqnarray} \label{10}
T_{BH}\!&=&\!\frac{V'(r_{+})\sqrt{-f(r_{+})}r_{+}}{8\pi
a(r_{+})}\nonumber\\\!&=&\!\frac{m[r_{+}^{2}(1-8j^{2}m-4jl)-2l^{2}]}{\pi
r_{+}^{2}\sqrt{-4j^{2}r_{+}^{6}+(1-8j^{2}m)r_{+}^{4}+2ml^{2}}},
\end{eqnarray}
where
\begin{eqnarray} \label{11}
V'(r_{+})=\partial _{r}V(r)|_{r=r_{+}}, \qquad f(r_{+})=1-\frac{2m}{r_{+}^{2}}, \qquad a(r_{+})=jr_{+}^{2}+\frac{ml}{r_{+}^{2}}.
\end{eqnarray}
The angular velocity at the event horizon is
\begin{eqnarray} \label{12}
\Omega_{+}=-\frac{f(r_{+})}{a(r_{+})}.
\end{eqnarray}
Now, applying the modified Dirac equation based on GUP and Kernerand Mann's fermion tunneling method, we focus on investigating the quantum correction to fermion tunneling from the Kerr-G\"{o}del black hole. In order to avoid the dragging effect, we first introduce a dragging coordinate transformation
\begin{eqnarray} \label{13}
{\rm d}\phi=\Omega {\rm d}t=\frac{4a(r)f(r)}{r^{2}V(r)-4a^{2}(r)}{\rm d}t.
\end{eqnarray}
Then, the Kerr-G\"{o}del black hole's metric can be expressed as
\begin{eqnarray} \label{14}
{\rm d}s^{2}= \!&-&
\!\frac{4a(r)f(r)}{r^{2}V(r)-4a^{2}(r)}{\rm d}t^{2}+\frac{1}{V(r)}{\rm d}r^{2}+\frac{r^{2}}{4}{\rm d}\theta^{2}\nonumber\\
\!&+&
\!\frac{r^{2}f(r)\sin^{2}\theta+(r^{2}V(r)-4a^{2}(r))\cos^{2}\theta}{4f(r)}{\rm d}\psi^{2}.
\end{eqnarray}

Here, we will only analysis the spin-up case since the final result is the same as the spin-down case as can be presented by using the methods described below. As for the spin-up case, we adopt the following ansatz
\begin{eqnarray} \label{15}
\psi_{\uparrow}(t,r,\theta,\psi)=\left(\begin{array}{clr} A(t,r,\theta,\psi) \\ 0
\\ B(t,r,\theta,\psi) \\ 0
\end{array}\right)\exp\left(\frac{i}{\hbar}I_{\uparrow}(t,r,\theta,\psi)\right),
\end{eqnarray}
and construct the gamma matrices of the G\"{o}del black hole as following
\begin{eqnarray} \label{16}
&&\gamma^{\,t}=\sqrt{-g^{tt}} \left(\begin{array}{clr} i &\,\,\, 0 \\
0 & \,\,\,i
\end{array}\right), \qquad \gamma^{\,r}=\sqrt{V(r)} \left(\begin{array}{clr} 0 & \sigma^{3} \\
\sigma^{3} & 0
\end{array}\right)\nonumber\\
&&\gamma^{\,\theta}=\sqrt{g^{\theta\theta}} \left(\begin{array}{clr} 0 & \sigma^{1} \\
\sigma^{1} & 0
\end{array}\right)\qquad
\gamma^{\,\psi}=\sqrt{g^{\psi\psi}} \left(\begin{array}{clr} 0 & \sigma^{2} \\
\sigma^{2} & 0
\end{array}\right)
\end{eqnarray}
and $ \sigma^{i}(i=1,2,3) $ are Pauli matrices as follows:
\begin{eqnarray} \label{17}
\sigma^{1}=\left(\begin{array}{clr} 0 & \,\,\,1
\\ 1 & \,\,\,0
\end{array}\right),\,\,\,\,\,\,\sigma^{2}=\left(\begin{array}{clr} 0 & -i
\\ i & \,\,\,0
\end{array}\right),\,\,\,\,\,\,\sigma^{3}=\left(\begin{array}{clr} 1
&\,\,\,
 0
\\ 0 & -1
\end{array}\right).
\end{eqnarray}

Substituting the ansatz (15) and gamma matrices (16) into the generalized Dirac equation (5), and adopting WKB approximation, after neglecting the high order of $\hbar$, we get the following four equations:
\begin{eqnarray} \label{18}
&&-iA\sqrt{-g^{tt}}\partial _{t}I_{\uparrow}-B\sqrt{g^{rr}}(1-\beta \mu^{2})\partial _{r}I_{\uparrow}-A\mu\beta\left[g^{rr}(\partial _{r}I_{\uparrow})^{2}+g^{\theta\theta}(\partial _{\theta}I_{\uparrow})^{2}+g^{\psi\psi}(\partial _{\psi}I_{\uparrow})^{2}\right]{}\nonumber\\{}&&+B\beta\sqrt{g^{rr}}\left[g^{rr}(\partial _{r}I_{\uparrow})^{2}+g^{\theta\theta}(\partial _{\theta}I_{\uparrow})^{2}+g^{\psi\psi}(\partial _{\psi}I_{\uparrow})^{2}\right]\partial _{r}I_{\uparrow}+A\mu(1-\beta\mu^{2})=0,
\end{eqnarray}
\begin{eqnarray} \label{19}
&&iB\sqrt{-g^{tt}}\partial _{t}I_{\uparrow}-A\sqrt{g^{rr}}(1-\beta \mu^{2})\partial _{r}I_{\uparrow}-B\mu\beta\left[g^{rr}(\partial _{r}I_{\uparrow})^{2}+g^{\theta\theta}(\partial _{\theta}I_{\uparrow})^{2}+g^{\psi\psi}(\partial _{\psi}I_{\uparrow})^{2}\right]{}\nonumber\\{}&&+A\beta\sqrt{g^{rr}}\left[g^{rr}(\partial _{r}I_{\uparrow})^{2}+g^{\theta\theta}(\partial _{\theta}I_{\uparrow})^{2}+g^{\psi\psi}(\partial _{\psi}I_{\uparrow})^{2}\right]\partial _{r}I_{\uparrow}+B\mu(1-\beta\mu^{2})=0,
\end{eqnarray}
\begin{eqnarray} \label{20}
&&A\big\{-(1-\beta\mu^{2})\sqrt{g^{\theta\theta}}\partial _{\theta}I_{\uparrow}+\beta\sqrt{g^{\theta\theta}}\partial _{\theta}I_{\uparrow}\left[g^{rr}(\partial _{r}I_{\uparrow})^{2}+g^{\theta\theta}(\partial _{\theta}I_{\uparrow})^{2}+g^{\psi\psi}(\partial _{\psi}I_{\uparrow})^{2}\right]{}\nonumber\\{}&&-i(1-\beta\mu^{2})\sqrt{g^{\psi\psi}}\partial _{\psi}I_{\uparrow}+i\beta\sqrt{g^{\psi\psi}}\left[g^{rr}(\partial _{r}I_{\uparrow})^{2}+g^{\theta\theta}(\partial _{\theta}I_{\uparrow})^{2}+g^{\psi\psi}(\partial _{\psi}I_{\uparrow})^{2}\right]\big\}=0,
\end{eqnarray}
\begin{eqnarray} \label{21}
&&B\big\{-(1-\beta\mu^{2})\sqrt{g^{\theta\theta}}\partial _{\theta}I_{\uparrow}+\beta\sqrt{g^{\theta\theta}}\partial _{\theta}I_{\uparrow}\left[g^{rr}(\partial _{r}I_{\uparrow})^{2}+g^{\theta\theta}(\partial _{\theta}I_{\uparrow})^{2}+g^{\psi\psi}(\partial _{\psi}I_{\uparrow})^{2}\right]{}\nonumber\\{}&&-i(1-\beta\mu^{2})\sqrt{g^{\psi\psi}}\partial _{\psi}I_{\uparrow}+i\beta\sqrt{g^{\psi\psi}}\left[g^{rr}(\partial _{r}I_{\uparrow})^{2}+g^{\theta\theta}(\partial _{\theta}I_{\uparrow})^{2}+g^{\psi\psi}(\partial _{\psi}I_{\uparrow})^{2}\right]\big\}=0.
\end{eqnarray}
It is difficult to directly obtain the solution of the above equations. Due to there are three Killing vectors $(\partial/\partial_{t})^{\mu}$, $(\partial/\partial_{\phi})^{\mu}$ and $(\partial/\partial_{\psi})^{\mu}$ in the five dimensional Kerr-G\"{o}del black hole, so the action $I_{\uparrow}$ in the dragging coordinate frame can be decomposed as
\begin{eqnarray} \label{22}
I_{\uparrow}=-(\omega-J\Omega)t+L\psi+R(r,\theta),
\end{eqnarray}
here $J$ and $\omega$ are the tunneling particle's angular momentum and energy, respectively. From Eqs. (20) and (21), we can get
\begin{eqnarray} \label{23}
(\sqrt{g^{\theta\theta}}\partial _{\theta}I_{\uparrow}+i\sqrt{g^{\psi\psi}}\partial _{\psi}I_{\uparrow})\big\{\beta \left[g^{rr}(\partial _{r}I_{\uparrow})^{2}+g^{\theta\theta}(\partial _{\theta}I_{\uparrow})^{2}+g^{\psi\psi}(\partial _{\psi}I_{\uparrow})^{2}\right]-(1-\beta\mu^{2})\big\}=0.
\end{eqnarray}
Due to $\beta$ is a very small quantity, the second term in above equation is not zero. Thus, we can obtain
\begin{eqnarray} \label{24}
\sqrt{g^{\theta\theta}}\partial _{\theta}I_{\uparrow}+i\sqrt{g^{\psi\psi}}\partial _{\psi}I_{\uparrow}=0.
\end{eqnarray}
That means
\begin{eqnarray} \label{25}
g^{\theta\theta}(\partial _{\theta}I_{\uparrow})^{2}+g^{\psi\psi}(\partial _{\psi}I_{\uparrow})^{2}=0.
\end{eqnarray}
Substituting Eqs. (22) and (25) into Eqs. (18) and (19), and canceling $A$ and $B$, we obtain the differential equation
\begin{eqnarray} \label{26}
&&\beta^{2}\frac{r^{2}f(r)V^{4}(r)}{r^{2}V(r)-4a^{2}(r)}(\partial _{r}R)^{6}+\beta\frac{r^{2}f(r)V^{3}(r)}{r^{2}V(r)-4a^{2}(r)}(\mu^{2}\beta-2)(\partial _{r}R)^{4}{}\nonumber\\{}&&
+\frac{r^{2}f(r)V^{2}(r)}{r^{2}V(r)-4a^{2}(r)}\left[(1-\beta\mu^{2})^{2}+2\beta\mu^{2}(1-\beta\mu^{2})\right](\partial _{r}R)^{2}{}\nonumber\\{}&&
-\frac{r^{2}f(r)V(r)}{r^{2}V(r)-4a^{2}(r)}\mu^{2}(1-\beta\mu^{2})^{2}-(\omega-J\Omega)^{2}=0.
\end{eqnarray}
From Eq. (26), we can see that $R(r,\theta)$ has explicit $r$-dependence and is independent of $\theta$. Thus, $R(r,\theta)=R(r)+\Theta(\theta)$ and $\partial _{r}R(r,\theta)=\partial _{r}R(r)$. Integrating at the event horizon $r_{+}$ and keeping the leading order term of $\beta$, we obtain the expression of the action $R(r)$
\begin{eqnarray} \label{27}
R_{\pm}(r)&=&\pm\int \frac{{\rm d}r}{rV(r)}\sqrt{\frac{r^{2}V(r)-4a^{2}(r)}{f(r)}\left[\mu^{2}\frac{r^{2}f(r)V(r)}{r^{2}V(r)-4a^{2}(r)}+(\omega-J\Omega)^{2}\right]}{}\nonumber\\{}
&&\times \left[1+\beta\mu^{2}+\beta \frac{r^{2}V(r)-4a^{2}(r)}{r^{2}f(r)V(r)}(\omega-J\Omega)^{2}\right]
{}\nonumber\\{}
&=&\pm i \pi \frac{(\omega-J\Omega_{+})r_{+}^{2}a^{2}(r_{+})}{\sqrt{-f(r_{+})}(r_{+}^{2}-r_{-}^{2})}(1+\beta\eta)
+\mbox{Real Part},
\end{eqnarray}
where $\pm$ correspond to the outgoing/ingoing solution, and
\begin{eqnarray} \label{28}
\eta=\frac{3}{2}\mu^{2}-2J\frac{\omega-J\Omega_{+}}{a(r_{+})}+\frac{\omega^{2}-J^{2}\Omega_{+}^{2}}{2f(r_{+})}
+\frac{(\omega-J\Omega_{+})^{2}[(r_{+}^{2}-r_{-}^{2})^{2}+4a^{2}(r_{+})(3r_{+}^{2}-r_{-}^{2})]}{f(r_{+})(r_{+}^{2}-r_{-}^{2})^{2}}.\nonumber\\
\end{eqnarray}
As Akhmedov E T, Akhmedova V et al. have indicated in Refs. \cite{51,52,53,54,55}, the expression of the tunneling rate $\Gamma\propto \exp \left(-\mbox{Im}\oint P\,{\rm d}r\right)$, instead of $\Gamma\propto \exp \left(-2\mbox{Im}\int P\,{\rm d}r\right)$ is invariant under canonical transformations. So the fermion's tunneling rate of the five-dimensional G\"{o}del black hole can be written as
\begin{eqnarray} \label{29}
\Gamma\propto &&\exp \left(-\mbox{Im}\oint P\,{\rm d}r\right){}\nonumber\\{}&&=\exp\left[-\mbox{Im}\left(\int P_{\text{out}}\,{\rm d}r-\int P_{\,\text{in}}\,{\rm d}r\right)\right]{}\nonumber\\{}&&=\exp\left[\mp 2\mbox{Im}\int P_{\text{out},\,\text{in}}\,{\rm d}r\right],
\end{eqnarray}
where $P=\partial_{r}R$, $\pm$ signs in front of this integral in last step correspond to $P_{\text{out}}$ and $P_{\,\text{in}}$, which have the relationship $P_{\text{out}}=-P_{\text{\,in}}$. So the tunneling probability of the five-dimensional Kerr-G\"{o}del black hole can be calculated as
\begin{eqnarray} \label{30}
\Gamma\propto &&\exp\left[-2\pi \frac{(\omega-J\Omega_{+})r_{+}^{2}a(r_{+})}{\sqrt{-f(r_{+})}(r_{+}^{2}-r_{-}^{2})}(1+\beta\eta)\right].
\end{eqnarray}
As the authors suggested in Refs. [51, 54,55], by calculating the tunneling rate in this way, the contribution from the time part of the action was overlooked. In order to work out the correct tunneling rate, we introduce general Kruskal coordinates $(T, R)$. Order
\begin{eqnarray} \label{31}
{\rm d}r^{*}=\sqrt{\frac{r^{2}V(r)-4a^{2}(r)}{4a(r)f(r)V(r)}}\,{\rm d}r.
\end{eqnarray}
As a result, we can write (14) as
\begin{eqnarray} \label{32}
{\rm d}s^{2}=&&\pm \frac{4a(r)f(r)}{r^{2}V(r)-4a^{2}(r)}\kappa^{-2}e^{-2\kappa r^{*}} \left(-{\rm d}T^{2}+{\rm d}R^{2}\right)+\frac{r^{2}}{4}{\rm d}\theta^{2}{}\nonumber\\{}&&+\frac{r^{2}f(r)\sin^{2}\theta+[r^{2}V(r)-4a^{2}(r)]\cos^{2}\theta}{4f(r)}{\rm d}\psi^{2},
\end{eqnarray}
where $+(-)$ sign stands for the Kerr-G\"{o}del metric outside (inside) the event horizon, and $\kappa=2\pi/T_{BH}$ is the Kerr-G\"{o}del black hole's surface gravity.
The region exterior of the Kerr-G\"{o}del black hole is
\begin{eqnarray} \label{33}
&&T_{\text{out}}=e^{\kappa r^{*}}\sinh (\kappa t),\nonumber\\
&&R_{\text{out}}=e^{\kappa r^{*}}\cosh (\kappa t),
\end{eqnarray}
where $\kappa=2\pi/T_{BH}$ is the surface gravity. The interior region is expressed as
\begin{eqnarray} \label{34}
&&T_{\text{in}}=e^{\kappa r^{*}}\cosh (\kappa t),\nonumber\\
&&R_{\text{in}}=e^{\kappa r^{*}}\sinh (\kappa t).
\end{eqnarray}

We can refer to [47-49], setting the time $t$ as $t\rightarrow t-\pi i/2\kappa$ an additional contribution coming from the time part of the action can be discoved that is $\mbox{Im}[(\omega-J\Omega_{+})\varDelta t_{\text{out,\,in}}]=-\pi (\omega-J\Omega_{+})/2\kappa$. Eventually, incorporating the temporal contribution, we turns out to be of the form
\begin{eqnarray} \label{35}
\Gamma\propto &&\exp\left\{\left[\mbox{Im}((\omega-J\Omega_{+})\varDelta t_{\text{out}})+\mbox{Im}((\omega-J\Omega_{+})\varDelta t_{\text{in}})-\mbox{Im}\oint P\,{\rm d}r\right]\right\}{}\nonumber\\{}&&
=\exp \left[-4\pi \frac{r_{+}^{2}a(r_{+})(\omega-J\Omega_{+})}{\sqrt{-f(r_{+})}(r_{+}^{2}-r_{-}^{2})}(1+\frac{1}{2}\beta\eta)\right].
\end{eqnarray}
As a consequence, the temperature including the modification terms of the G\"{o}del black hole can be determined as
\begin{eqnarray} \label{36}
T=\frac{\sqrt{-f(r_{+})}(r_{+}^{2}-r_{-}^{2})}{4\pi r_{+}^{2}a(r_{+})(1+\beta\eta/2)}=T_{BH}(1-\frac{1}{2}\beta\eta),
\end{eqnarray}
where
\begin{eqnarray} \label{37}
T_{BH}=\frac{\sqrt{-f(r_{+})}(r_{+}^{2}-r_{-}^{2})}{4\pi r_{+}^{2}a(r_{+})}=\frac{r_{+}V'(r_{+})\sqrt{-f(r_{+})}}{8\pi a(r_{+})}
\end{eqnarray}
is the Kerr-G\"{o}del black hole's Hawking temperature. According to the above expressions (35) and (36), we can see that nonlinear terms arise. That is to say, when the quantum gravity effect is taken into account, the emission spectrum of the higher dimensional G\"{o}del black hole is no more pure thermal. This means the information can be supplied. And the revised tunneling temperature is no longer the traditional Hawking temperature but depends on the emitted particle's mass $\mu$, angular momentum $J$ and energy $\omega$. If gravity corrected term is neglected, the results are consistent with the previous those in the literatures \cite{14,17}.

\section{Conclusion}
Comparing with three or four dimensional spacetime, five dimensional case becomes more complicated. As mentioned former, it is significate to investigate quantum correction to higher dimensional case. In this current paper, we have studied the fermion tunneling from a five dimensional Kerr G\"{o}del black hole when incorporating quantum gravity effects. For this purpose, we have first introduced a dragging coordinate transformation. For a rotating black hole, due to there exist dragging effect, a reasonable tunneling radiation should be described in a dragging coordinate framework. Here we have made the coordinate transformation $\phi=\Omega dt$, departing from one in Refs. [28-30], where the angular is fixed when integrating the action for the rotating spacetime. This transformation is more natural and convenient to derive tunneling radiation, especially for the higher dimensional case, we can get exact expression of the tunneling probability and tunneling temperature. After making the dragging coordinate transformation, applying Kerner and Mann's fermion tunneling method and the spacetime's symmetries of the Kerr G\"{o}del black hole, we have resolved the modified Dirac equation proposed by authors in Refs. [28-30]. As a result, the revised tunneling probability has been workout, and the corrected tunneling temperature also discovered. We find out that tunneling rate and tunneling temperature depend on the tunneling particle's mass, energy and angular momentum. Furthermore, there exists nonlinear terms in the imaginary part of the action. This means tunneling rate is no longer purely thermal and the information may be supplied in correlations between tunneling modes.

\acknowledgments

This work is supported in part by the Science and Technology Project Foundation of the Education Department of Liaoning Province, China (Grant No. L2011195).

\end{document}